# Spatially Periodic Cells Are Neither Formed From Grids Nor Poor Isolation

J. Krupic[1], N. Burgess[2], J. O'Keefe[1, 3]

[1]Research Department of Cell and Developmental Biology, UCL, Gower Street, London WC1E 6BT, UK

[2]Institute of Neurology, UCL, Queen Square, London WC1N 3BG, UK; Institute of Cognitive Neuroscience, UCL, Queen Square, London WC1N 3AR, UK.

[3]Sainsbury Wellcome Centre, University College London, London WC1E 6BT, UK.

**Grid cells recorded in the parahippocampal formation of freely moving rodents provide a strikingly periodic representation of self-location whose underlying mechanism has been the subject of intense interest. Our previous work(*1*) showed that grid cells represent the most stable subset of a larger continuum of spatially periodic cells (SPCs) which deviate from the hexagonal symmetry observed in grid cells. Recently Navratilova et al(*2*) suggested that our findings reflected poor isolation of the spikes from multiple grid cells, rather than the existence of actual non-grid SPCs. Here we refute this suggestion by showing that: (i) most SPCs cannot be formed from hexagonal grids; (ii) all standard cluster isolation measures are similar between recorded grid cells and non-grid SPCs, and are comparable to those reported in other laboratories; (iii) the spikes from different fields of band-like SPCs do not differ. Thus the theoretical implications of the presence of cells with spatially periodic firing patterns that diverge from perfect hexagonality need to be taken seriously, rather than 'explained away' on the basis of hopeful but unjustified assumptions.**

We recently reported neuronal firing in the medial entorhinal cortex and adjacent parasubiculum of freely moving rats whose spatial distribution across the environment was periodic and could be modelled as a superposition of periodic bands(*1*). We termed these "spatially periodic cells" (SPCs), of which about 37% were grid cells, having regular triangular grid-like firing patterns(*3*). We suggested that grid cells present the most stable subclass of spatially periodic cells which generally can deviate from the hexagonal symmetry (e.g. by becoming less regular, more elliptical, or more band-like). As noted by Navratilova et al(*2*) the presence of SPCs whose firing patterns are not perfectly translation invariant is theoretically important because they are incompatible with the current continuous attractor model of grid cell firing(*4–7*).

Navratilova et al(*2*) propose that all spatially periodic cells are actually grid cells, and that the appearance of non-grid SPCs reflects poor isolation of action potentials ("spikes") from multiple grid cells in our extracellular recordings. Our recordings were made using tetrodes(*8–11*). This is now a standard method used in hundreds of studies, including almost all of those reporting grid cells. The standard criteria used to measure isolation quality of the tetrode-recorded waveform clusters associated with each single unit are the L ratio and isolation distance(*12*), as well as the refractory period which indicates that no single unit should show spikes occurring within 2 ms of each other.

Navratilova et al(*2*) suggest the grid-like firing patterns of multiple local grid cells could sum to produce the appearance of periodic bands if their spatial offsets are approximately equal to 50% of grid wavelength. The authors suggest that the net firing of local grid cell clusters might have been identified as single SPCs by simple misclassification. The proposed 'contamination hypothesis' is refuted by examples of recorded non-grid SPCs where the pattern cannot result from any simple summation of offset grid cells. Such an example SPC is shown in figure 1, along with a simultaneously recorded grid cell. All SPCs are shown in Fig. S3 of our original paper(*1*). This cell exemplifies the problem: any single grid cell cannot account for all the observed fields: missing some fields and predicting additional fields which were not observed. Adding further grid cells merely results in the addition of further firing fields that were not observed.

Navratilova et al.'s(*2*) hypothesis predicts that different firing fields in SPCs actually come from different cells. This hypothesis can be tested by comparing the similarity of waveforms corresponding to each field (figure 2), similar to Fyhn et al.'s(*13*) procedure of showing the similarity of the waveforms corresponding to each grid cell field (Fig. S3(*13*)). The overlap in cluster space of spikes from different 'fields' suggests that they belong to the same cell.

Navratilova et al.'s(*2*) hypothesis further predicts that, because non-grid SPCs must be a summation of several grid cells, whereas grid cells could be single or multiple units, the mean firing rate of non-grid SPCs should be higher than those of grid cells, and the cluster isolation should be lower. As we had clearly shown (Fig. S8B(*1*)), the mean firing rate for non-grid SPCs was lower than that for grid cells (1.04±0.08 Hz versus 1.21 ± 0.11 Hz), but not significantly so (P=0.18, $t_{242}$=1.33). Furthermore, it was comparable to the mean firing rates previously reported for grid cells(*14–16*). We also note that there were no significant differences between peak firing rates or spike widths between grid cells and non-grid SPCs (original paper Fig. S8(*1*)). Equally, as we had clearly shown in our original paper (Fig. S8C-D(*1*)), cluster isolation did not differ between grid cells and non-grid SPCs: the mean L ratio was 0.016±0.003 vs. 0.017±0.002 (P=0.82, $t_{241}$=0.24) and the mean isolation distance was 39.9±14.7 vs. 36.8±9.6(P=0.85, $t_{241}$=0.19). In addition, the cluster quality as well as the overall percentage of hexagonal grid cells we found in mEC was comparable to those reported by other laboratories routinely recording grid cells (*14–16*). Furthermore, there was no significant correlation between isolation and gridness score (figure 6; L ratio vs gridness: P=0.85 and ρ=-0.012; isolation distance vs gridness: P=0.79 and ρ=-0.018). Finally, there was no significant difference in refractory periods between grid cells and non-grid SPCs (P = 0.95, t = -0.06, df = 242; two-sample t-test) the majority of which had an inter-spike >2 ms. Seven grid cells (out of 91) and four non-grid SPCs (out of 153) had a few spikes within the range of 1.5-1.95 ms (see Data Table 1), none of these cells were identified as a band-like cells (see Data Table2).

Although many types of divergence from hexagonal regularity were reported in our paper, Navratilova et al(*2*) concentrate on examining the validity of band-like SPCs in particular, as these firing patterns could potentially result from a superposition of grids. Like all SPCs, the firing of these cells show multiple sub-peaks consistent with a superposition of periodic bands, we did not report finding the constituent perfectly uniform bands themselves. In order to ensure that the band-like firing pattern did not result from combining several offset grids we compared the similarity of waveforms corresponding to each 'field' within a band (figures 3-4). Again, the overlap in cluster space of spikes from different 'fields' suggests that they belong to the same cell. In addition we re-analysed the assignment of spikes to single units using an automated cluster-kwik procedure(*17*) (figure 5; the raw data from these cells can be freely downloaded from 'space-memory-navigation.org'), an automated cutting technique routinely used by other laboratories(*18–20*). Our manually cut units coincided with the automatically separated units (mean±s.d.

overlap: 93±9%). The number of spikes in automatically cut clusters was (mean±s.d.) 44±22% larger than the number of spikes in manually cut clusters possibly due to an extra effort by us to make sure that no false positive spikes were included in the analysis. It also can be clearly seen from the clusters that the recorded triggering spike amplitudes are considerably larger than a detection threshold. Hence it is unlikely that spikes would be missed out due to a high detection threshold, thus the 'missing fields' in Figure 1 could not be statistical 'false negatives'.

Thus we welcome Navratilova et al's(*2*) request for more detail to reinforce our observation that non-grid SPCs exist, and are unlikely to be due to problems of isolating responses from an underlying population of perfect grid cells. The irregularity of the firing patterns of SPCs were sometimes related to geometrical layout of the environment, and spontaneous deviations from hexagonality were also observed, as with elliptical grid cells, observations that have been replicated by our and other laboratories(*21–26*). In conclusion, future work might usefully consider the theoretical implications of the presence of cells with spatially period firing patterns that diverge from perfect hexagonality, rather than attempting to explain them away on the basis of speculations and simulations.

## Acknowledgments

The research was supported by grants from the Wellcome Trust and the Gatsby Charitable Foundation. J.K. is a Wellcome Trust Sir Henry Welcome Fellow. This work was conducted in accordance with the UK Animals (Scientific Procedures) Act (1986). All the methods are described in (*1*).

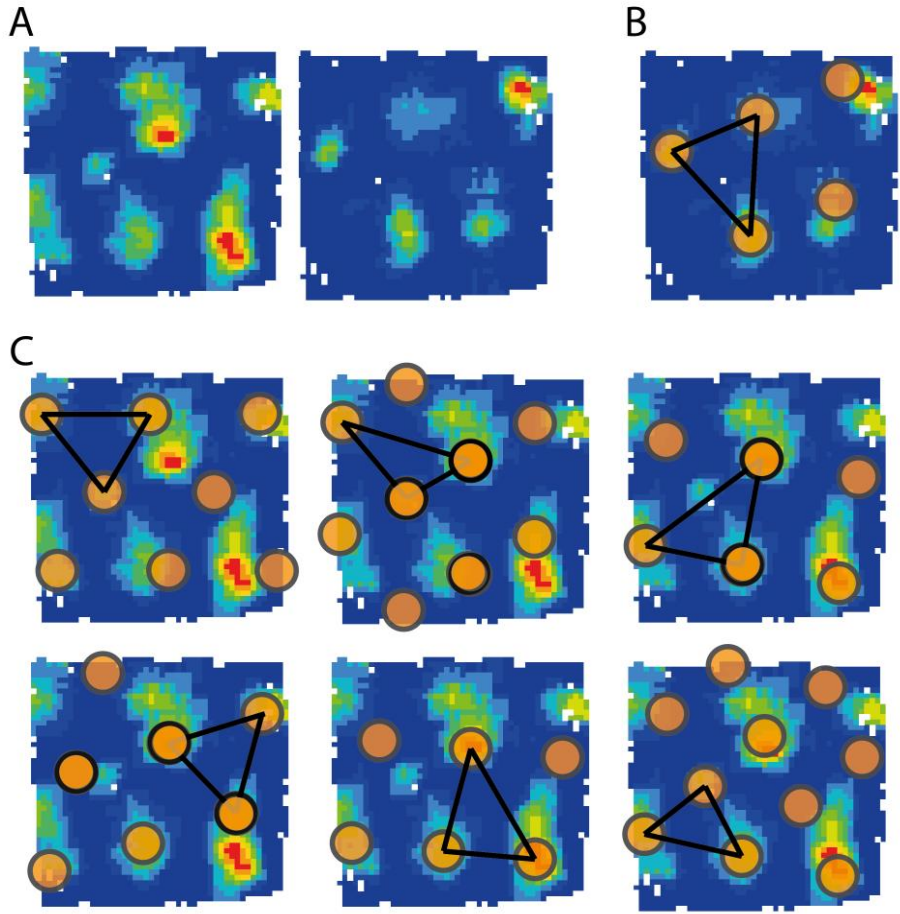

**Figure 1:** An example of non-grid SPC which cannot be 'generated' by a simple summation of several grid cells and a simultaneously recorded grid cell (rat r1738). **A:** Example rate maps of a spatially periodic non-grid cell (left) and a co-recorded grid cell (right). **B:** The grid cell can be accurately approximated by a single elliptical unit grid (superimposed orange circles). **C:** No unit grid can well approximate the firing rate map of the spatially periodic non-grid cell with some fields missing and importantly some new fields appearing.

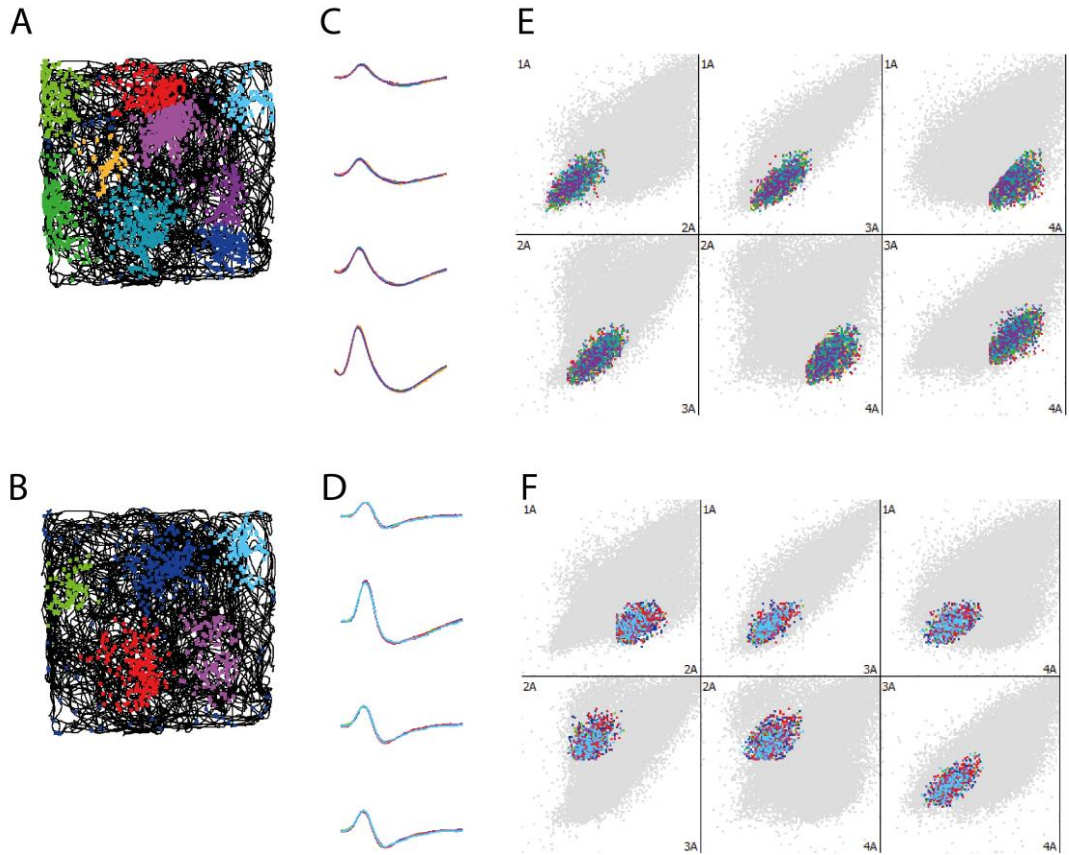

**Figure 2:** Rate maps of cell in fig.1 cut into individual 'sub-fields' (different colours represent spikes recorded in distinct 'sub-fields'; rat trajectory shown in black). **A-B**: rate maps of a spatially periodic non-grid cell and a grid cell cut into individual 'sub-fields' (different colours represent spikes recorded in distinct 'sub-fields'; rat trajectory shown in black) with their mean waveforms superimposed **(C-D)** showing complete overlap in cluster space **(E-F)** suggesting that all recorded spikes come from the same single unit.

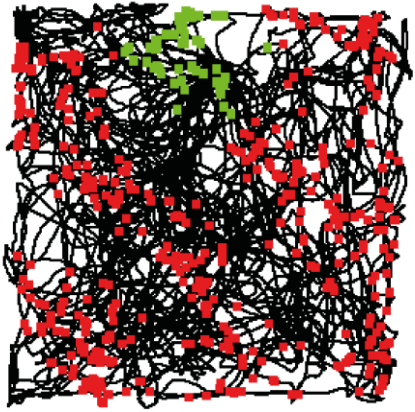 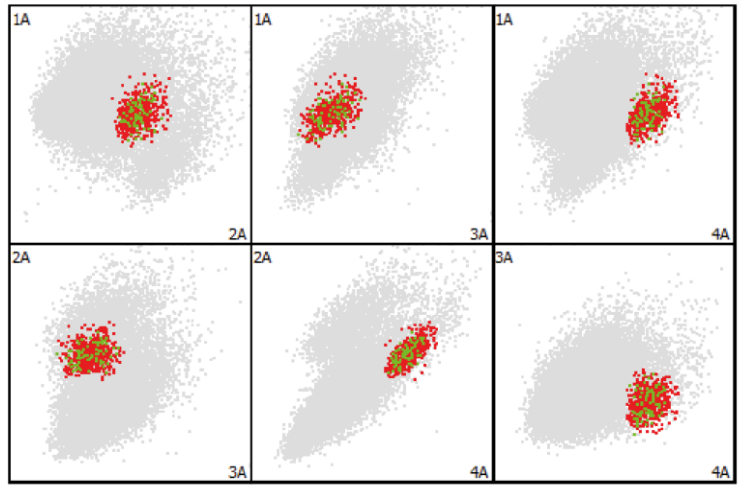

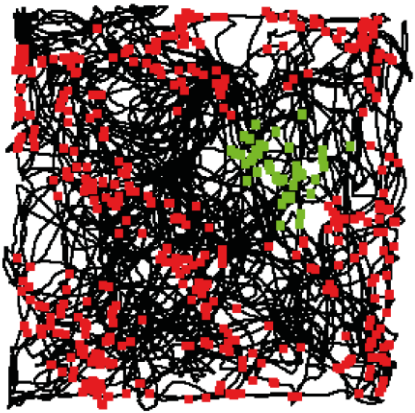 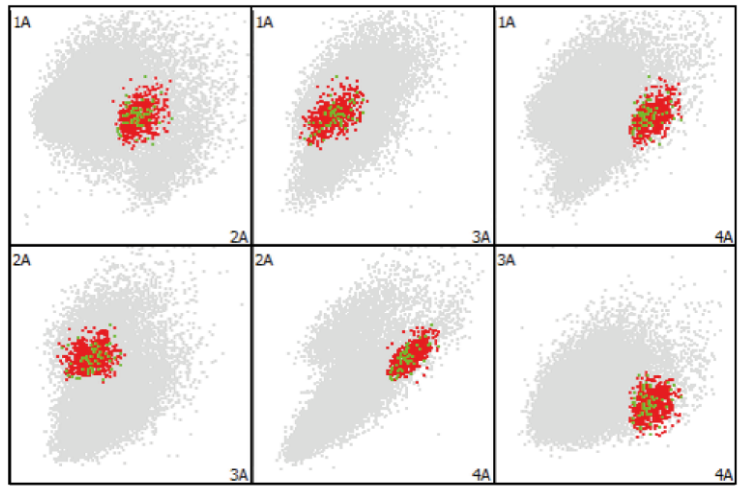

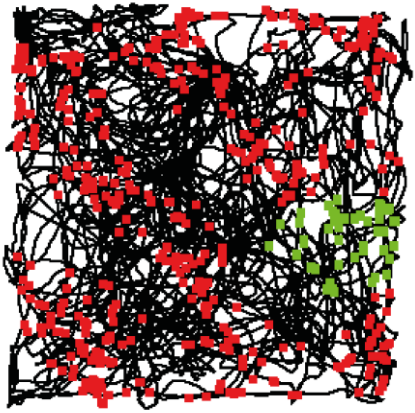 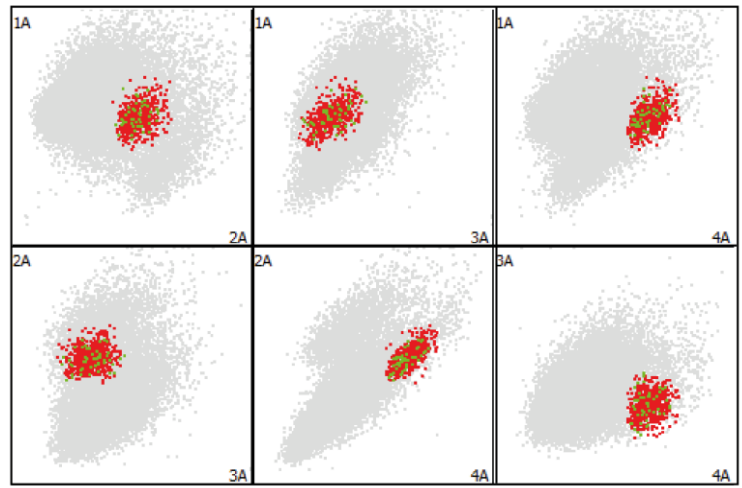

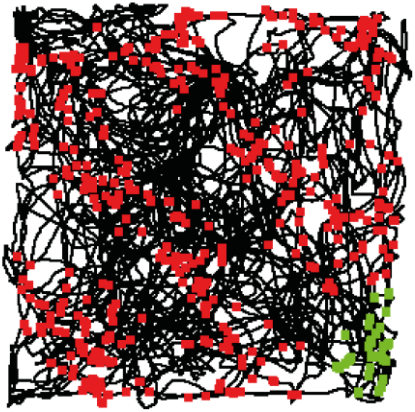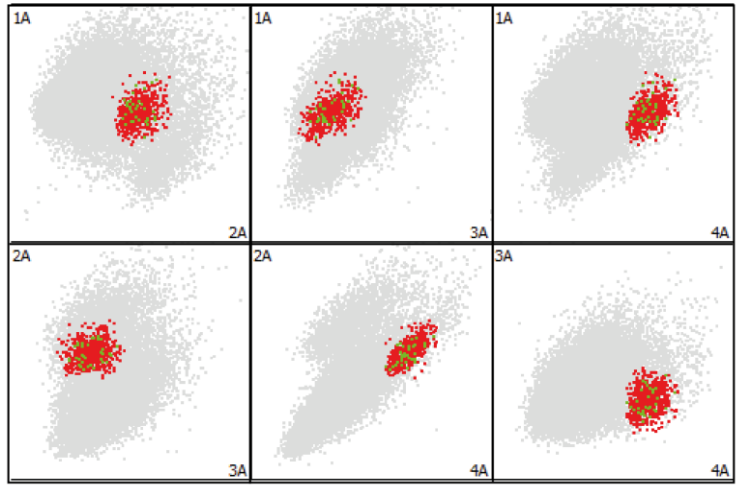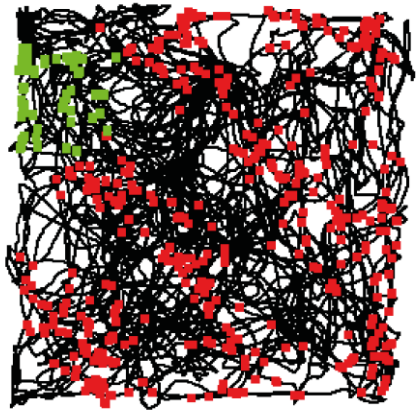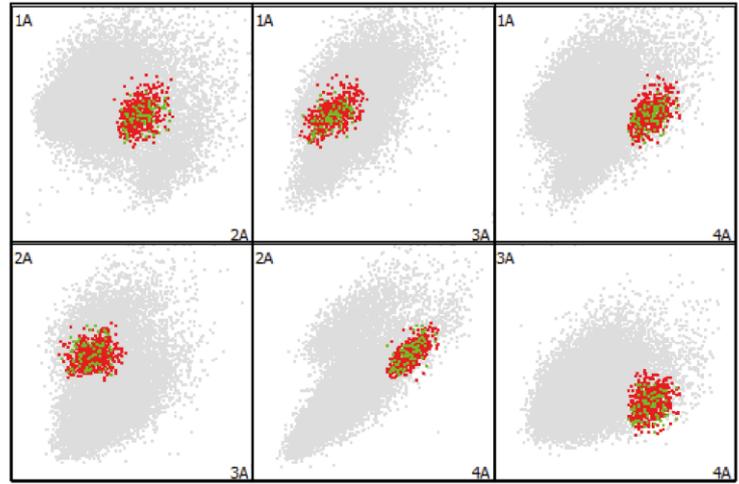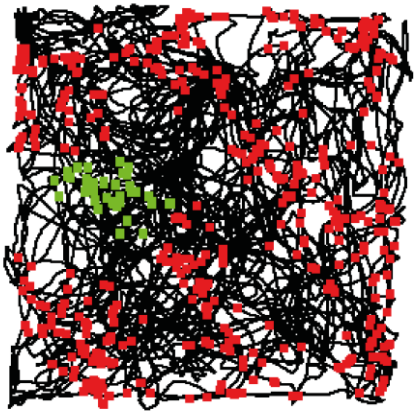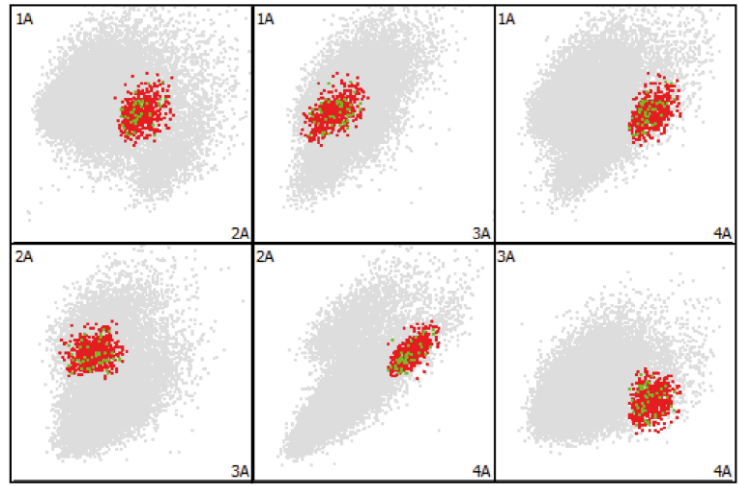

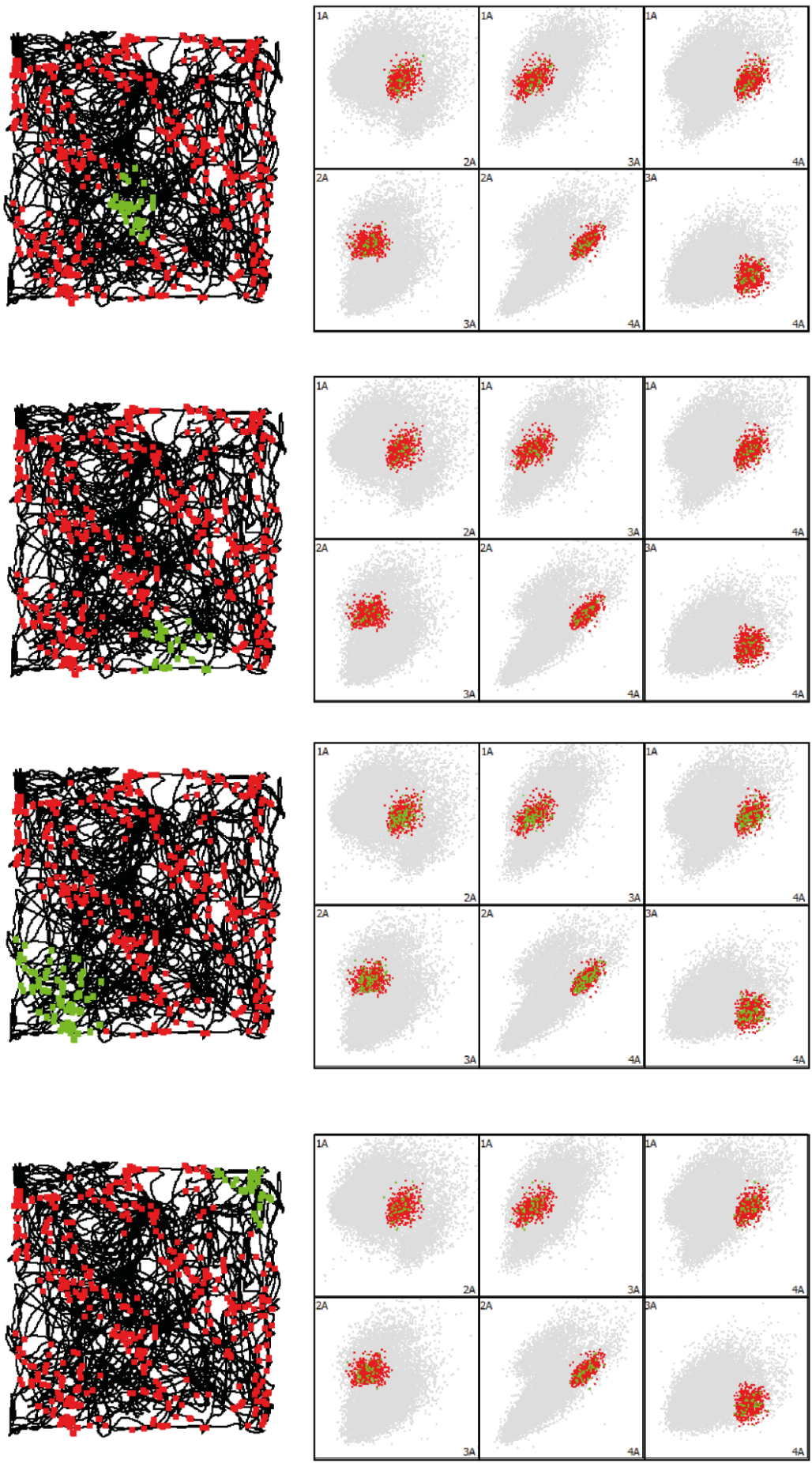

**Figure 3:** A typical example of spikes from a band-like cell (red) exhaustively cut into individual 'sub-fields' (green) which completely overlap in cluster space (right) suggesting that all recorded spikes belong to the same single unit.

A

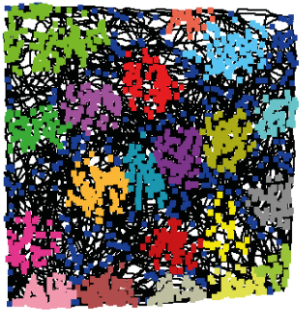 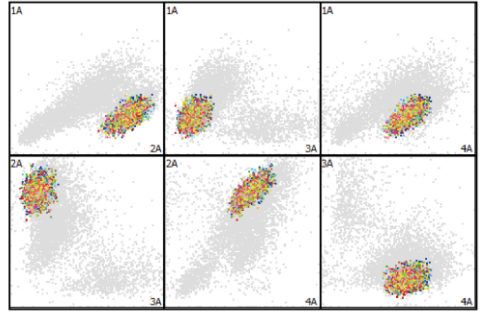 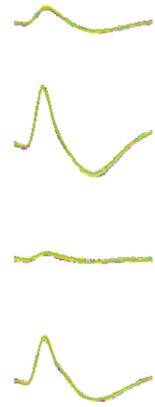

B

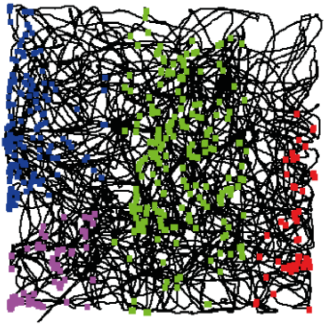 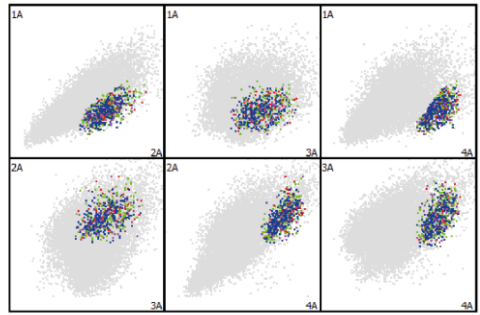 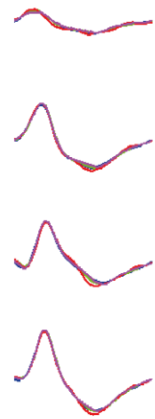

C

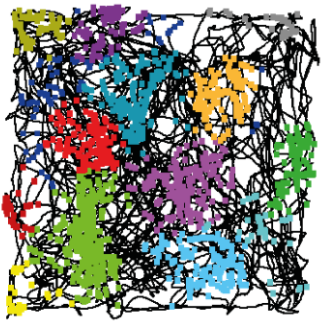 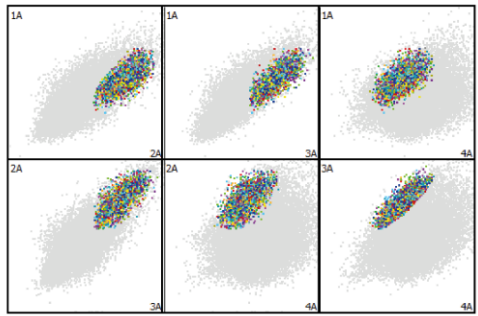 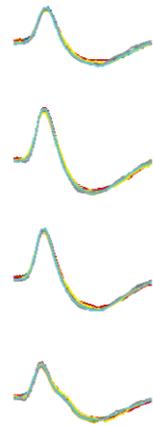

D

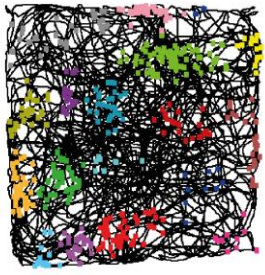 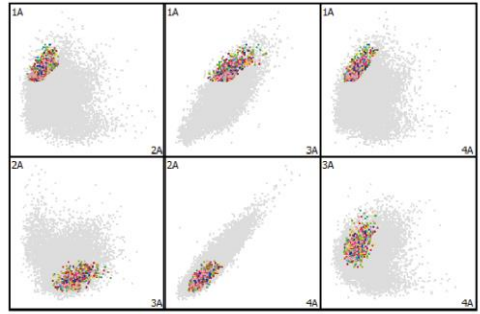 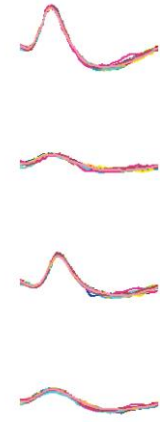

E

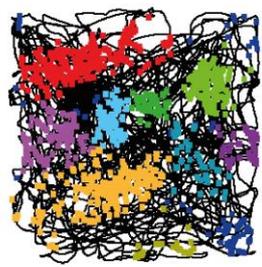 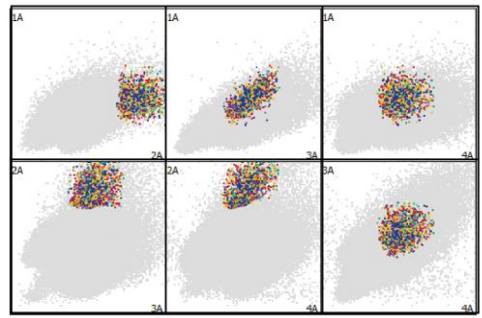 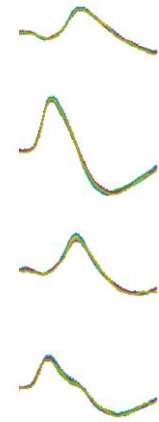

F

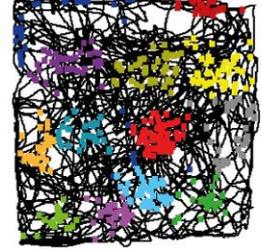 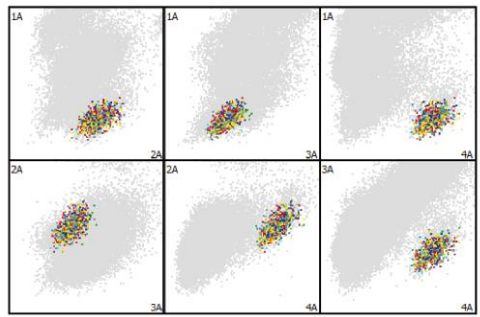 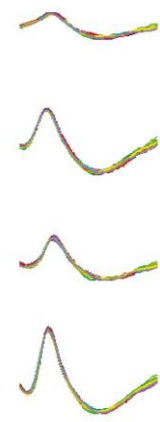

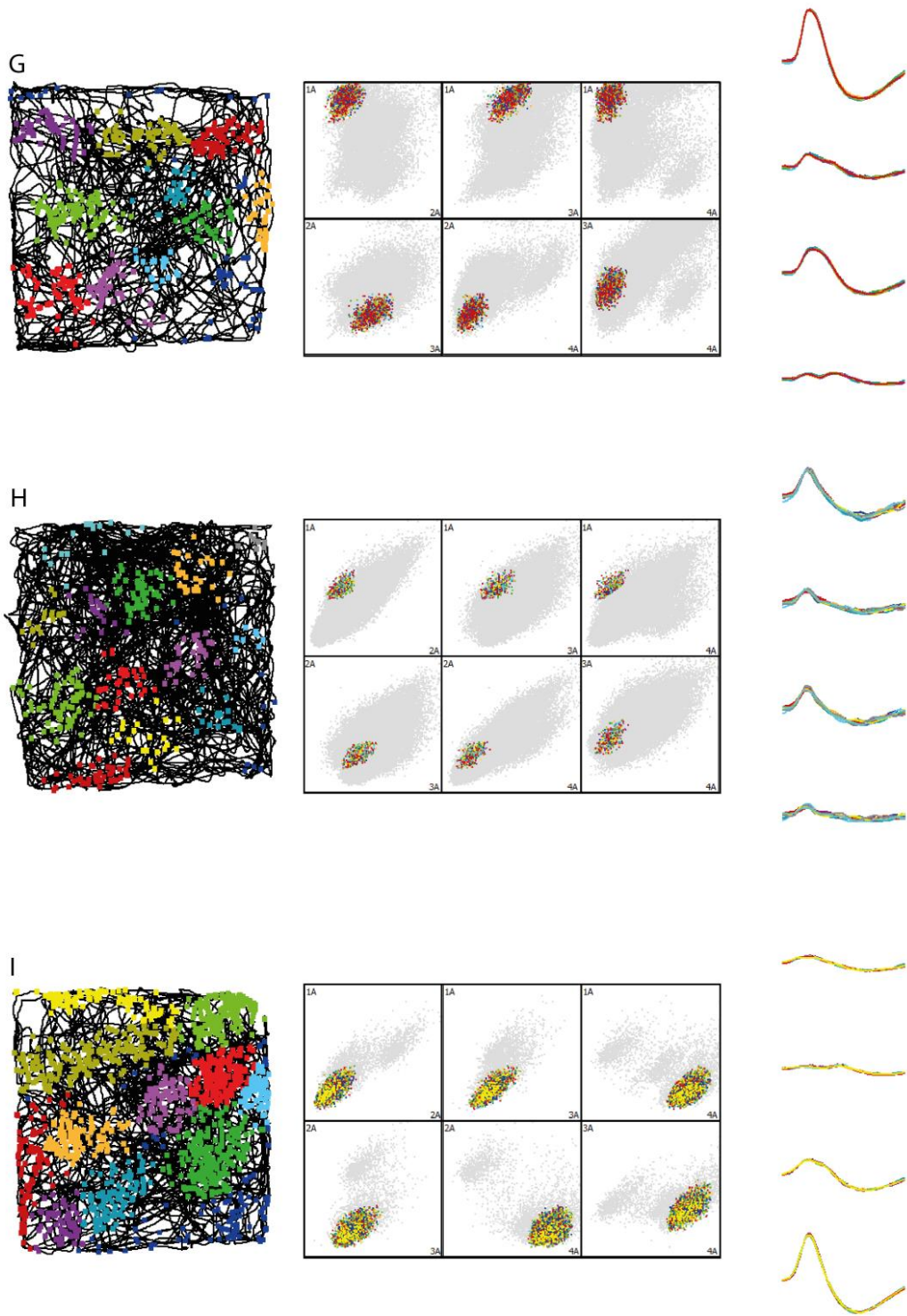

**Figure 4:** All band-like cells **(A-I)**, except the cell shown in figure 3, cut into individual 'sub-fields' (left, different colours represent spikes from distinct 'fields'; rat's trajectory shown in black) with their mean waveforms superimposed (right) showing a complete overlap in a cluster space (middle) suggesting that all recorded spikes come from the same single unit.

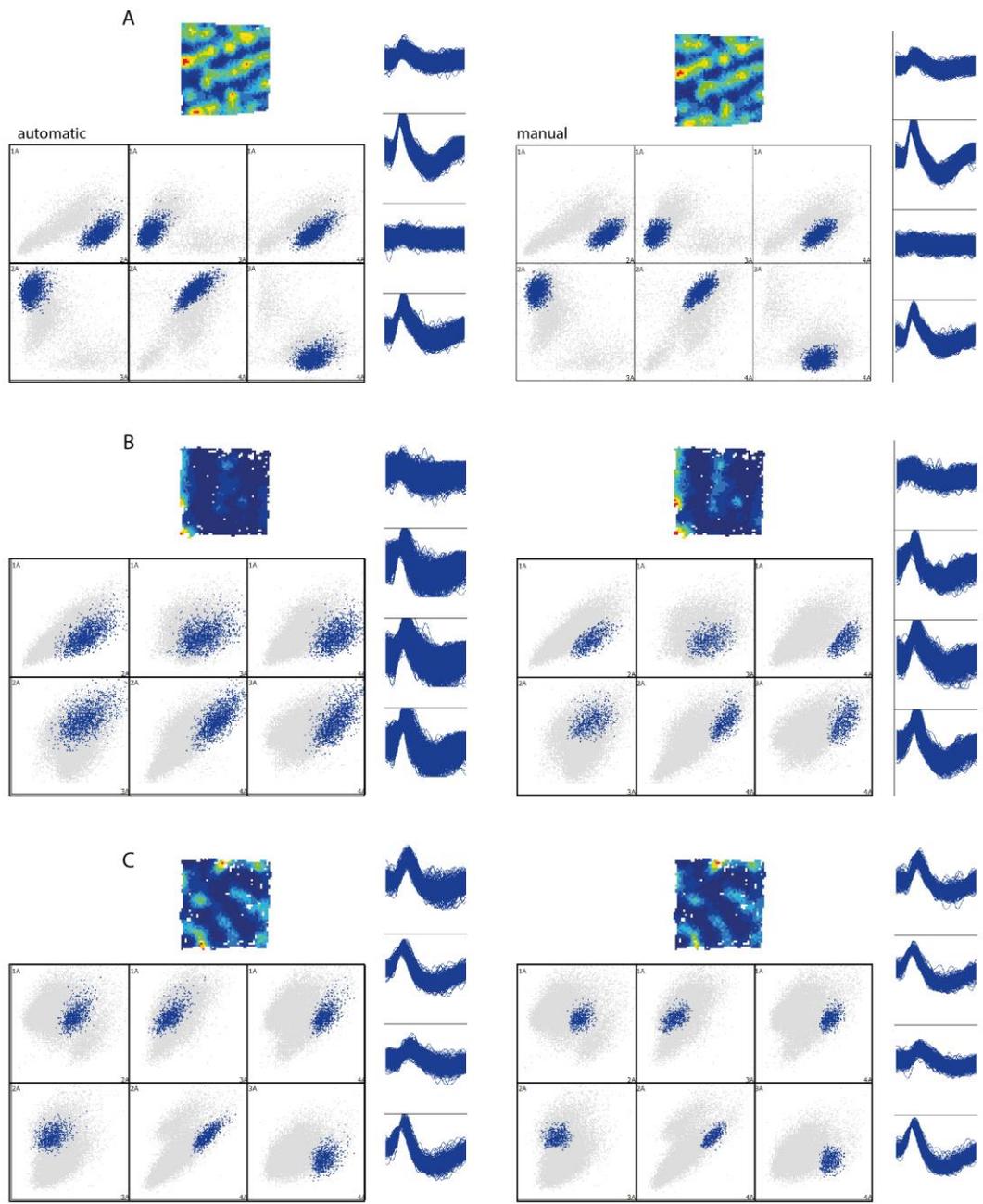

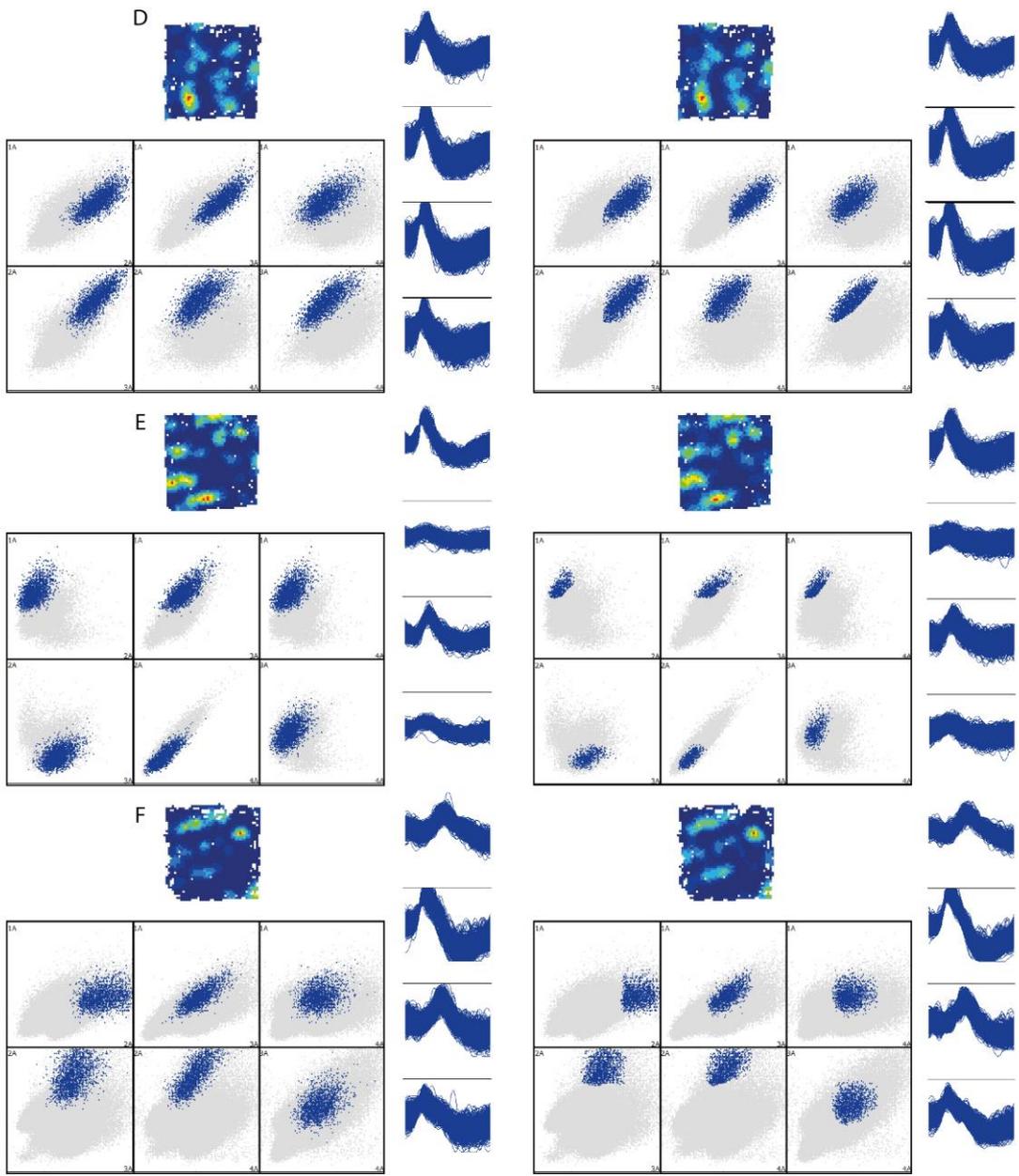

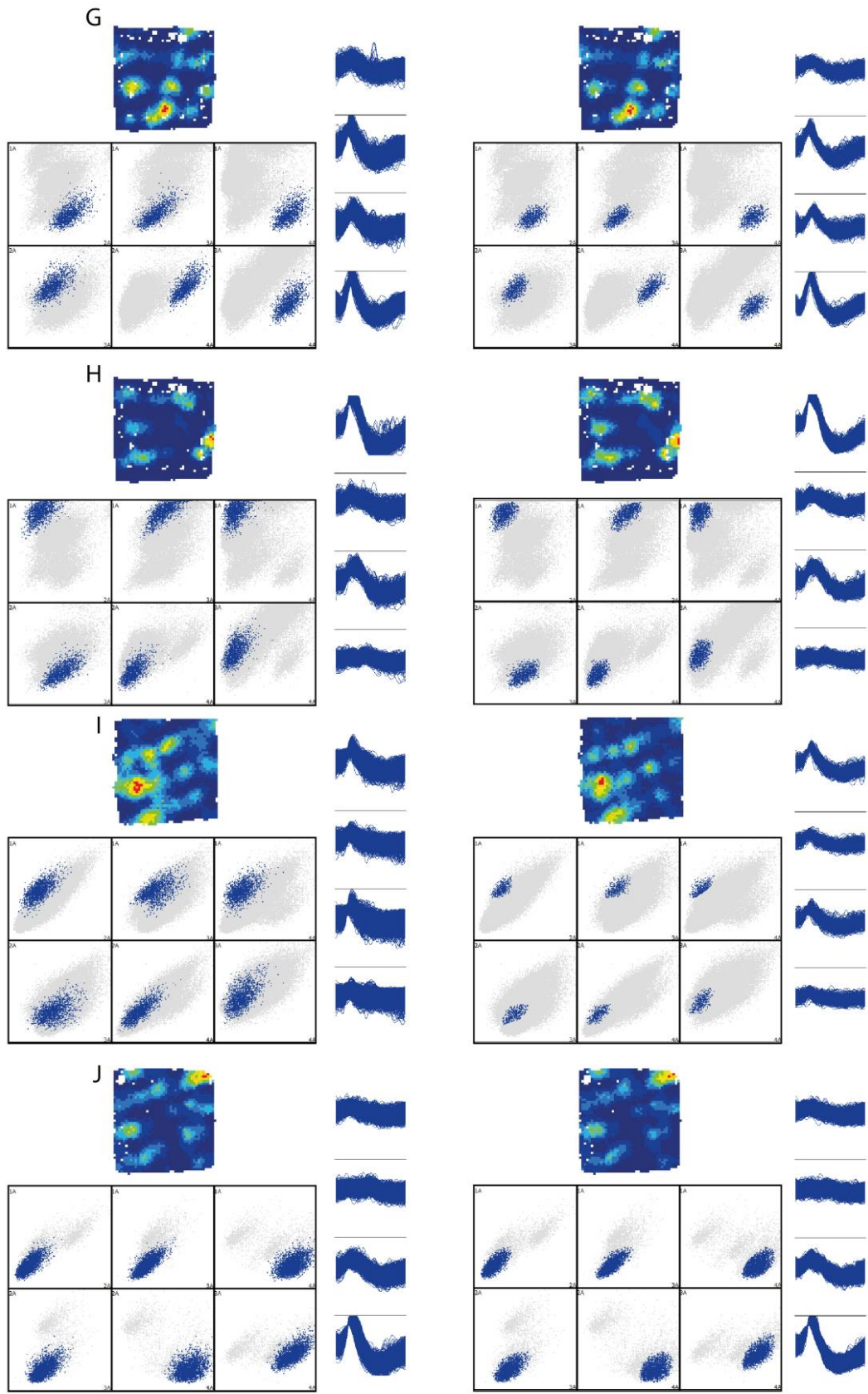

**Figure 5:** All band-like cells (A-J) cut using the automatic kwik-cluster algorithm (left) vs. cut manually (right). Rate maps (top), isolated cells and their waveforms shown in blue together with all other recorded spikes (grey).

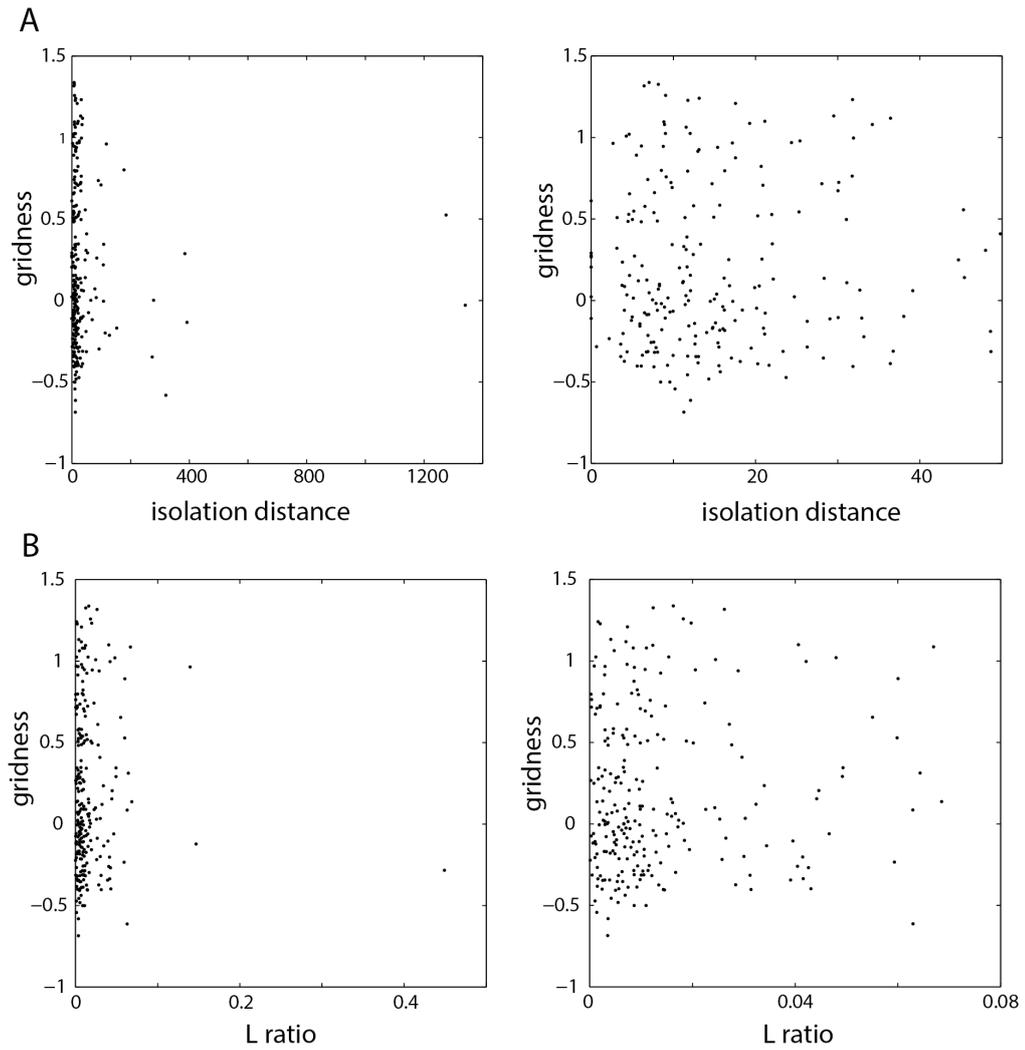

**Figure 6:** Gridness vs. cluster isolation. Absence of a significant correlation ρ between the gridness and isolation distance **(A)** and L ratio **(B)**. Right figures show magnifications of the leftmost portion of the figures on the left.

**Data Table 1.** Spatially periodic cells with at least one inter-spike interval <2 ms. (* indicates grid cells).

| Rat number | gridness | N spikes | Trial duration, s | Min interspike interval, ms | N spikes < 2ms | L ratio | Isolation distance |
|---|---|---|---|---|---|---|---|
| r1738 | 0.09 | 1776 | 1801 | 1.8 | 4 | 0.010 | 16.8 |
| r1738 | 0.53* | 5464 | 1800 | 1.9 | 5 | 0.007 | 22.0 |
| r1710 | 1.33* | 337 | 900 | 1.6 | 1 | 0.012 | 8.1 |
| r1710 | 1.32* | 467 | 900 | 1.5 | 1 | 0.026 | 6.4 |
| r1728 | -0.13 | 671 | 900 | 1.5 | 1 | 0.005 | 26.3 |
| r1728 | -0.21 | 529 | 901 | 1.9 | 1 | 0.001 | 21.1 |
| r1728 | 0.58* | 454 | 901 | 1.8 | 1 | 0.005 | 15.6 |
| r1728 | 0.41* | 1257 | 900 | 1.9 | 1 | 0.03 | 49.8 |
| r1728 | 0.66* | 329 | 900 | 1.8 | 1 | 0.012 | 7.7 |
| r1737 | -0.29 | 894 | 900 | 1.9 | 1 | 0.009 | 15.5 |
| r1737 | 1.21* | 2115 | 1200 | 1.8 | 2 | 0.007 | 17.6 |

**Data Table 2**. Inter-spike intervals of band-like cells.

| Rat number | N spikes | Trial duration, s | Min interspike interval, ms |
|---|---|---|---|
| r1710 | 1872 | 1200 | 4.9 |
| r1737 | 629 | 901 | 2.7 |
| r1738 | 571 | 900 | 3.0 |
| r1738 | 2013 | 900 | 2.1 |
| r1682 | 625 | 1201 | 3.9 |
| r1682 | 1446 | 901 | 2.3 |
| r1682 | 652 | 901 | 2.8 |
| r1682 | 997 | 901 | 2.0 |
| r1682 | 367 | 1201 | 8.9 |
| r1682 | 2463 | 1200 | 2.1 |